# Introduction of Probability Density Alternation Method for Inverse Analyses of Integral Equations in Surface Science

Keito Hashidate[1], Rieko Iwayasu[1], Takumi Otake[1], and Ken-ichi Amano[2,(a)*]

[1] Graduate School of Agriculture, Meijo University, Nagoya 468-8502, Japan

[2] Faculty of Agriculture, Meijo University, Nagoya 468-8502, Japan

[(a)] E-mail: amanok@meijo-u.ac.jp

**Abstract** — Integral equations frequently arise in surface science, and in some cases, they must be treated as inverse problems. In our previous work on optical tweezers, atomic force microscopy, and surface force measurement apparatus, we performed inverse calculations to obtain the pressure between parallel plates from measured interaction forces. These inverse analyses were used to reconstruct solvation structures near solid surfaces and density distribution profiles of colloidal particles. In the course of these studies, we developed a method that enables inverse analyses through a unified and systematic procedure, hereafter referred to as the Probability Density Alternation (PDA) method. The central idea of this method is to reformulate a given integral equation in terms of probability density functions. In this letter, we demonstrate the validity of the PDA method both analytically and numerically. While the PDA method is less advantageous for single integral equations, it becomes a convenient and powerful approach for inverse analyses involving double or higher-order integral equations.

**Introduction.** — Integral equations appear in a wide range of scientific fields. Examples include scattering problems of electromagnetic waves [1,2] and ultrasound [3], image reconstruction in astronomy [4], force detection in mechanical engineering [5,6], and analysis in chemical physics [7]. Our research focuses on surface physical chemistry, and the goal is to understand the density profile of colloidal particles [8-13] and the viscosity distribution of a liquid [14] near a surface, and the pair potential between two substances [15]. In studies of density profiles, inverse analyses of integral equations are required, in which the pressure between two parallel flat plates is calculated inversely from interactions measured by optical tweezers (OT) [10], atomic force microscopy (AFM) [9,11,12], and surface force apparatus (SFA) [8,13]. Once the pressure between the flat plates is obtained through the inverse analysis, a density distribution profile of colloidal particles near a substrate can be reconstructed.

In the course of these studies, we developed an inverse analysis method that applies probability density function (PDF), which we refer to as the Probability Density Alternation (PDA) method. The advantages of the PDA method are that it enables an inverse analysis through a systematic procedure, and that it allows double or higher-order integral equations to be treated with relative ease. As a simple and intuitive explanation, we illustrate the PDA method using a single integral equation

$$f(s) = \int_{x_a}^{x_b} g(h(x;s))u(x)dx, \tag{1}$$

where $f$, $h$, and $u$ are known functions, $g$ is an unknown function, $s$ is an arbitrary parameter, and the integral range is from $x_a$ to $x_b$. This integral equation can be seen as $(x_b - x_a)$ times an average height of the integrand from the viewpoint of Monte Carlo integration. Then, we replace the integral in eq. (1) with an integral that calculates the average height of $g(h(x;s))u(x)$ from $x_a$ to $x_b$:

$$f(s) = (x_b - x_a)\int_{h_a}^{h_b} P_s(h)\bar{u}_s(h)g(h)dh, \tag{2}$$

where $P_s(h)$ is the PDF of $h$, which depends on $s$. When the integration variable $x$ is changed to $h$, $u(x)$

must be converted to a function of $h$: $\bar{u}_s(h)$. Since $h$ is the function of $x$ and $s$, $x$ can be transformed to a function of $h$ and $s$. Hence, if $h(x;s)$ is an injective function for $x \in [x_a, x_b]$, $u(x)$ can be transformed to $\bar{u}_s(h)$. The lower limit $h_a$ and the upper limit $h_b$ are respectively the minimum and maximum of $h(x;s)$ in the considered numerical range. The form of eq. (2) is a reformulation of eq. (1) in terms of PDF, Monte Carlo integration, and Lebesgue integration. In the PDA method, the original integral equation is transformed into a common integral equation with a kernel function $M_s$:

$$f(s) = \int_{h_a}^{h_b} M_s(h) g(h) dh. \tag{3}$$

In the equation, $M_s$ is expressed in terms of the remaining functions excluding $g(h)$. This integral equation can be readily transformed into a matrix form:

$$\boldsymbol{f} = \boldsymbol{M_s g}. \tag{4}$$

Therefore, the unknown vector $\boldsymbol{g}$ (*i.e.*, $g(h)$) can be obtained by using the inverse matrix of $\boldsymbol{M_s}$,

$$\boldsymbol{g} = \boldsymbol{M_s}^{-1} \boldsymbol{f}. \tag{5}$$

In some cases, it can be solved by additionally applying the determinant-optimization method. For example, the determinant-optimization method is important in an inverse scattering problem [2], where the sensor positions are purposely designed to optimize the determinant of the matrix.

In this letter, we first analytically demonstrate the validity of the PDA method by comparing the PDA-based solution with our previously established general solution for the single integral equation in the OT system [10]. Next, assuming a probe tip with a flattened apex (*i.e.*, blunt tip), we present the PDA-based solution of a single integral equation in the AFM system. The difficulty of this equation lies in the presence of the unknown function both inside and outside the integral. We refer to the PDA method adapted to this condition as the δ-PDA method. The complete agreement between the solution from a general process and the solution obtained by the δ-PDA method provides an analytical proof of the validity of the δ-PDA method. Finally, for a double integral equation in the SFA system [13], we numerically demonstrate the correctness and practical utility of the PDA method.

**PDA method for a single-integral equation in OT.** — Here, we prove the validity of the PDA method by showing that the general solution we have previously presented matches the solution obtained by the PDA method. The OT system is shown in fig. 1, where two large particles 1 and 2 are trapped by OT. The large particles with radius $r_L$ are surrounded by the small particles with radius $r_S$. The distance between the centers of the two large particles is denoted by $s$, and the distance between the centers of the two contacting small particles located at $\theta$ is denoted by $h$. In this study, our aim is to obtain the pressure between the flat plates ($\phi$). This is because, the pressure is an important input for calculation of the small particle's density profile around the large particle. Although we do not calculate the density profile in the present study, the detailed calculation method has been described in our past paper [9]. The definition of $h$ is somewhat complicated. However, we intentionally defined it in order to obtain the pressure induced by the dispersing small particles. The space into which the centers of the small particles cannot enter corresponds to the excluded volume of the large particle, and $r$ represents the radius of the excluded volume. In our paper [9], we introduced a single integral equation to inversely obtain the pressure between the flat plates from the force between the large particles ($f$):

$$f(s) = 2\pi r^2 \int_0^{\pi/2} \phi(h) \sin\theta \cos\theta \, d\theta. \tag{6}$$

The variable $h$ is a function of $s$ and $\theta$, $h = s - 2r\sin\theta$, which makes the inverse calculation slightly difficult, because it cannot readily be changed to a matrix form. However, eq. (6) can be transformed into a solvable

form through several derivations as follows [9]

$$\frac{f(s)}{\pi} = \int_{s-2r}^{s} (s-h)\phi(h)dh. \qquad (7)$$

The above equation can be transformed to the matrix form and $\phi(h)$ can be obtained using the inverse matrix.

In what follows, we convert eq. (6) into eq. (7) using the PDA method to demonstrate the method's validity. From fig.1, the following relationships can be written: $\sin\theta = (s - h)/(2r)$; $\cos\theta = [1 - (s - h)/(2r)]^{1/2}$. Introducing the following definition, $\sin\theta\cos\theta \equiv u(\theta)$, it can be rewritten as a function of $s$ and $h$ as follows:

$$\bar{u}_s(h) = \frac{s-h}{2r}\sqrt{1 - \left(\frac{s-h}{2r}\right)^2} \qquad (8)$$

Since $h = s - 2r\sin\theta$ and $\theta \in [0, \pi/2]$, the PDF of $h$ being $P_s(h)$ is

$$P_s(h) = 1\bigg/\left(\pi r\sqrt{1 - \left(\frac{s-h}{2r}\right)^2}\right), \qquad (9)$$

where $h \in [s - 2r, s]$ and $P_s(h)$ is zero when $h \notin [s - 2r, s]$. We set $s$ to $s \in [s_a, s_b]$, and using $\theta \in [0, \pi/2]$, the minimum ($h_a$) and maximum ($h_b$) of $h$ can be expressed as follows:

$$h_a = \min_{s \in [s_a, s_b], \theta \in [0, \pi/2]} \{s - 2r\sin\theta\} = s_a - 2r; \qquad (10)$$

$$h_b = \max_{s \in [s_a, s_b], \theta \in [0, \pi/2]} \{s - 2r\sin\theta\} = s_b. \qquad (11)$$

Then, in the PDA method, eq. (6) can be given by

$$\frac{f(s)}{2\pi r^2} = \frac{\pi}{2}\int_{h_a}^{h_b} P_s(h)\bar{u}_s(h)\phi(h)dh. \qquad (12)$$

We define a following equation $P_s(h)\bar{u}_s(h) \equiv M_s(h)$, leading to a solvable integral form. The solvable form can be changed to the matrix form, and it can be analysed inversely using the inverse matrix. However, to proof the PDA method, we should continue the explanation. Considering $h \in [s - 2r, s]$, the integration can be partitioned as follows:

$$\frac{f(s)}{\pi^2 r^2} = \int_{h_a}^{s-2r} M_s(h)\phi(h)dh + \int_{s-2r}^{s} M_s(h)\phi(h)dh + \int_{s}^{h_b} M_s(h)\phi(h)dh. \qquad (13)$$

Since $P_s(h) = 0$ in the range $h \notin [s - 2r, s]$, $M_s(h) = 0$ in the same range also. Thus, only the second integral in eq. (13) remains. Finally, substituting eq. (8) and eq. (9) into eq. (13), one can obtain exactly the same form as eq. (7). We successfully showed agreement between the integral form derived from the PDA method and that obtained by the general method [9], and the PDA method is thereby verified.

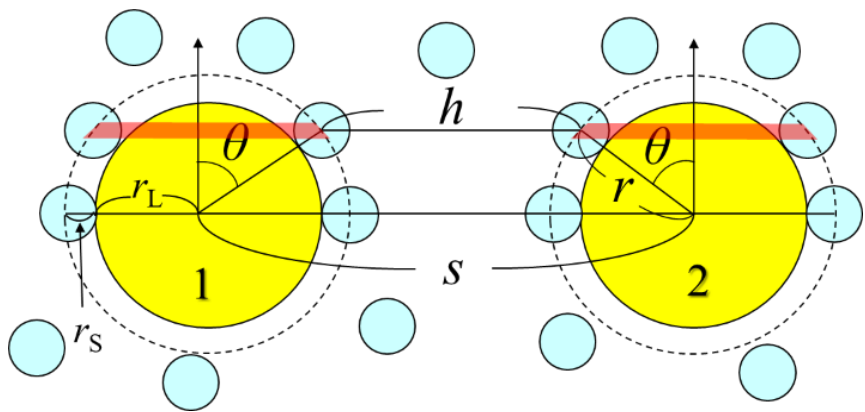

Fig. 1: Illustration of the OT system. Sum of the pressures between the red surface elements with distance $h$ corresponds to the force between the large particles 1 and 2.

**δ-PDA method for a single-integral equation in AFM.** — Here, we tackle a single integral equation having unknown functions inside and outside of the integral. To solve such an integral equation, we introduce a modified PDA method named δ-PDA method. We demonstrate the validity of the δ-PDA method by showing the solution from a general analytical method matches the solution obtained by the δ-PDA method. The AFM with a blunt tip is shown in fig. 2, where the large cut sphere with a flat bottom represents the blunt tip. The hemisphere like tip has radius $r_P$, and the radius of the small particle is $r_S$. The distance between the centers of the blunt tip and the contacting small particle is $r$, while the distance between the centers of the small particles contacting the blunt tip and the flat substrate is denoted by $h$. In this study, our aim is to obtain the pressure between the flat plates from the AFM force curve. This is because, the pressure is an important input for calculation of the colloidal particle's density profile on a flat substrate underneath. Although we do not calculate the density profile in the present study, the transform method from the pressure into the density profile has been described in our past paper [9,11,12]. The definition of $h$ for the AFM system is also somewhat complicated. However, we deliberately defined it in order to obtain the pressure induced by the dispersing colloidal particles. When the small particle is much smaller than the blunt tip ($r_P \gg r_S$), we can approximate $s' \approx h(\theta_{max})$. To obtain the density profile of the small particles near the substrate, the pressure between flat surfaces in the fluid ($\phi$) should be prepared beforehand [9,11,12]. For this purpose, we introduce a single integral equation which connects $\phi$ and force between the blunt tip and the substrate $f$:

$$f(s') = \phi(s')A_b + 2\pi r^2 \int_0^{\theta_{max}} \phi(h(\theta;s'))\sin\theta\cos\theta d\theta. \qquad (14)$$

Here, the area of the tip's flat bottom is denoted by $A_b$. Firstly, we show a solvable form obtained from a general derivation process (the detailed derivation is not shown but the hint is described in [9]):

$$\frac{f(s')}{2\pi} = \int_{s'}^{s'+r\sin\theta_{\max}} \left\{ (s' + r\sin\theta_{\max} - h) + \frac{A_b}{2\pi}\delta(h - s') \right\} \phi(h) dh. \tag{15}$$

The solvable form can be transformed to the matrix form, where the function on the left-hand side and $\phi$ are multidimensional vectors and the integrand excluding $\phi$ is the square matrix. In the square matrix, the delta function is Kronecker delta function. Therefore, the matrix equation can be analysed inversely using the inverse matrix.

In what follows, we use the δ-PDA method and show that the resultant integral equation is the same as eq. (14). That is, we convert eq. (14) into eq. (15) using the δ-PDA method to demonstrate the method's validity. As shown in fig. 2, $h = s - r_S - r\sin\theta_{\max}$ and $s' = s - r_S - r\sin\theta_{\max}$, and hence the following two expressions are obtained: $\sin\theta = (s' + r\sin\theta_{\max} - h)/r$; $\cos\theta = [1 - (s' + r\sin\theta_{\max} - h)/r]^{1/2}$. In addition, $\theta$ varies only within the first quadrant; in other words, $\sin\theta$ and $\cos\theta$ are always positive. Introducing the following definition, $\sin\theta\cos\theta \equiv u(\theta)$, it can be rewritten as a function of $s'$ and $h$ as follows:

$$\bar{u}_{s'}(h) = \frac{s' + r\sin\theta_{\max} - h}{r} \sqrt{1 - \left(\frac{s' + r\sin\theta_{\max} - h}{r}\right)^2}. \tag{16}$$

In the δ-PDA method, one must prepare the PDF of $h(\theta; s')$ within the integral range $\theta \in [0, \theta_{\max}]$, which can be given by

$$P_{s'}(h) = 1 \bigg/ \left( r\theta_{\max} \sqrt{1 - \left(\frac{s' + r\sin\theta_{\max} - h}{r}\right)^2} \right) \tag{17}$$

in the AFM system. We set the scan range of the approach/retract as $s' \in [0, s_{\max}]$, and using $\theta \in [0, \theta_{\max}]$, the minimum ($h_a$) and maximum ($h_b$) of $h$ are expressed as follows:

$$h_{\mathrm{a}} = \min_{\theta \in [0, \theta_{\max}], s' \in [0, s'_{\max}]} h(\theta; s'); \tag{18}$$

$$h_{\mathrm{b}} = \max_{\theta \in [0, \theta_{\max}], s' \in [0, s'_{\max}]} h(\theta; s'). \tag{19}$$

Based on the above considerations, the integral in eq. (14) can be transformed from the integral with respect to $\theta$ to one with respect to $h$,

$$f(s') - A_{\mathrm{b}}\phi(s') = (\theta_{\max} - 0)2\pi r^2 \int_{h_{\mathrm{a}}}^{h_{\mathrm{b}}} \phi(h) P_{s'}(h) \bar{u}_{s'}(h) dh. \tag{20}$$

We consider that $h_{\mathrm{a}} = s'$ and $h_{\mathrm{a}} \leq s' \leq h_{\mathrm{b}}$. The function $\phi(s')$ on the left-hand side in eq. (20) can be rewritten as

$$\phi(s') = \int_{h_{\mathrm{a}}}^{h_{\mathrm{b}}} \phi(h) \delta(h - s') dh, \tag{21}$$

where the delta function is a one-sided delta function. This integral equation is convenient, because $\phi(s')$ on the left-hand side in eq. (20) can be moved to the inside of the integral. Then, we finally obtain

$$\frac{f(s')}{2\pi} = \int_{h_{\mathrm{a}}}^{h_{\mathrm{b}}} \left\{ (s' + r\sin\theta_{\max} - h) + \frac{A_{\mathrm{b}}}{2\pi} \delta(h - s') \right\} \phi(h) dh, \tag{22}$$

where $h_{\mathrm{a}} = s'$ and $h_{\mathrm{b}} = s' + r\sin\theta_{\max}$. The forms of both eq. (15) and eq. (22) are the same. Therefore, validity of the $\delta$-PDA method was analytically verified.

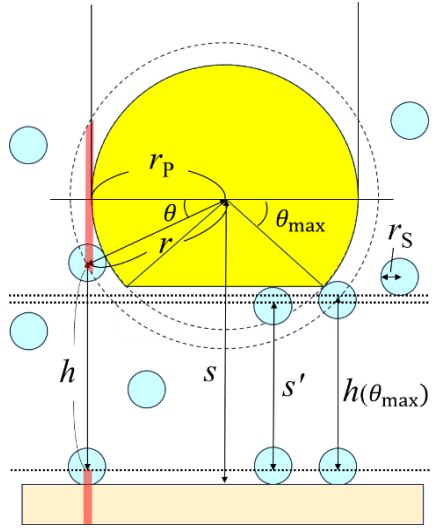

Fig .2: Illustration of the AFM system with a blunt tip. Sum of the pressures between the red surface elements with distance $h$ corresponds to the force between the blunt tip and the substrate.

**PDA method for a double-integral equation in SFA.** — Here, we demonstrate the correctness and practical utility of the PDA method by applying a double integral equation in the SFA system. The SFA system is shown in fig. 3, where the cylinders 1 and 2 are arranged orthogonally. They are immersed in a colloidal suspension of spherical small particles with radius $r_S$. The radius of each cylinder is $r_C$ and its length is much longer than $r_S$. The central axis of C1 is at $x = z = 0$ and the central axis of C2 is at $y = 0$ and $z = s + 2r$, where $r = r_S + r_C$ and $s$ is the closest distance between the excluded volume surfaces of the cylinders. Also, the distance between micro planar elements is denoted by $h$ (see fig. 3). In this study, our aim is to obtain the pressure between the flat plates. This is because, the pressure is an important input for calculation of the colloidal particle's density profile on the substrate. Although we do not calculate the density profile in the present study, the transform method from the pressure into the density profile has been described in our past paper [8,13]. The definition of $h$ in the SFA system is somewhat complicated. However, we deliberately defined it in order to obtain the pressure induced by the dispersing small particles. As shown in fig. 3, the force ($f$) between the crossed cylinders can be approximately expressed as the sum of the pressure ($\phi$) between the facing surface elements:

$$f(s) = 4\int_0^r \int_0^r \phi\left(s + 2r - \sqrt{r^2 - x^2} - \sqrt{r^2 - y^2}\right) dxdy. \tag{23}$$

The pressure function is expressed as $\phi(h)$ and the variable $h$ is written as $h = s + 2r - (r^2 - x^2)^{1/2} - (r^2 - y^2)^{1/2}$, which makes the inverse calculation difficult. However, eq. (23) can be transformed into a solvable form through somewhat complicated derivations as follows:

$$f(s) = 4\int_{s-r}^{s} \frac{s-w}{\sqrt{r^2 - (s-w)^2}} Q(w) dw; \tag{24}$$

$$Q(w) = \int_{w+r}^{w+2r} \frac{w + 2r - h}{\sqrt{r^2 - (w + 2r - h)^2}} \phi(h) dh. \tag{25}$$

The above equations have the kernel functions, and they can be alternated to the matrix forms. Then, $\phi(h)$ can be obtained by using the two inverse matrices of the kernel matrices. More specifically, $Q(w)$ is calculated from the matrix form of eq. (24) in the first half, and then $\phi(h)$ is obtained from the matrix form of eq. (25) in the latter half. Although the derivation process of eqs. (24) and (25) is not shown here, the process involves difficult derivations which requires somewhat brilliant ideas. In other words, it cannot be solved with a simple systematic process. Meanwhile, in the PDA method, one can solve eq. (23) with a simpler process. Moreover, only a single inverse matrix is used to solve. To demonstrate the PDA method, we convert eq. (23) into a new single integral equation with a kernel function. From eq. (23), the ranges of variables $x$ and $y$ are $x \in [0, r]$ and $y \in [0, r]$, respectively. The scan range is set as $s \in [s_a, s_b]$ and the range of $h$ is $h \in [s, s + 2r]$. In the PDA method, the PDF of $h$ being $P_s(h)$ can be prepared numerically. That is, $h$ can be seen as a function of variables $x$, $y$ and $s$ (i.e., $h$ is $h(x,y;s)$), which gives a histogram of $h$ (i.e., $P_s(h)$). Then, eq. (23) can be rewritten as

$$\frac{f(s)}{4r^2} = \int_{h_a}^{h_b} P_s(h)\phi(h) dh, \tag{26}$$

where

$$h_{\mathrm{a}} = \min_{s\in[s_{\mathrm{a}},s_{\mathrm{b}}], x\in[0,r], y\in[0,r]} \left\{s + 2r - \sqrt{r^2 - x^2} - \sqrt{r^2 - y^2}\right\} = s_{\mathrm{a}}; \tag{27}$$

$$h_{\mathrm{b}} = \max_{s\in[s_{\mathrm{a}},s_{\mathrm{b}}], x\in[0,r], y\in[0,r]} \left\{s + 2r - \sqrt{r^2 - x^2} - \sqrt{r^2 - y^2}\right\} = s_{\mathrm{b}} + 2r. \tag{28}$$

In eq. (26), $r^2$ originates from the integral domain of the double integral in eq. (23). Consequently, eq. (26) can be written by a matrix form with a single square matrix ($\boldsymbol{P}_s$) constructed by both $P_s(h)$ and $dh$. When $\boldsymbol{P}_s$ is the regular matrix, $\phi(h)$ can be obtained by using the inverse matrix as follows:

$$\boldsymbol{\phi} = \boldsymbol{P}_s^{-1} \boldsymbol{F}, \tag{29}$$

where $\boldsymbol{\phi}$ and $\boldsymbol{F}$ are the vectors of $\phi(h)$ and $f(s)/(4r^2)$, respectively.

In what follows, we show the validity of the PDA method (see fig. 4). The calculation condition in this figure is as follows: $r_{\mathrm{S}}$ = 50 nm; $r_{\mathrm{C}}$ = 1 cm; $s_{\mathrm{a}}$ = 0 nm; $s_{\mathrm{b}}$ = 1000 nm; the matrix size of $\boldsymbol{P}_s$ is 1000 × 1000. The thick blue curve is the inputted pressure between the flat walls (benchmark), which we arbitrary prepared. The red thin curve is the reconstructed result by eq. (29). As shown in the figure, the shapes of the two curves match well. This corroborates that the PDA method also functions well in the double-integral case. We also report that we have confirmed that accurate reconstructions are possible under the various pressure (benchmark) conditions. We note that if $x$ and $y$ exist in the same term in the function of $h$, that is, if $x$ and $y$ are in a multiplication or division relationship in $h$, it is very hard for a general inverse method to derive a matrix form with two regular matrices, but it is possible for the PDA method to derive a matrix form. Hence, the PDA method is considered to be a more universal inverse calculation technique. We note that when the double-integral equation includes a function $u(x,y)$ in the integrand, the PDA method slightly more difficult. However, it can be solved by modifying the double axes into the single axis. This is a straightforward technique for that case. However, when the shape of $u(x,y)$ is relatively simple or relatively flat, we can propose a more efficient technique.

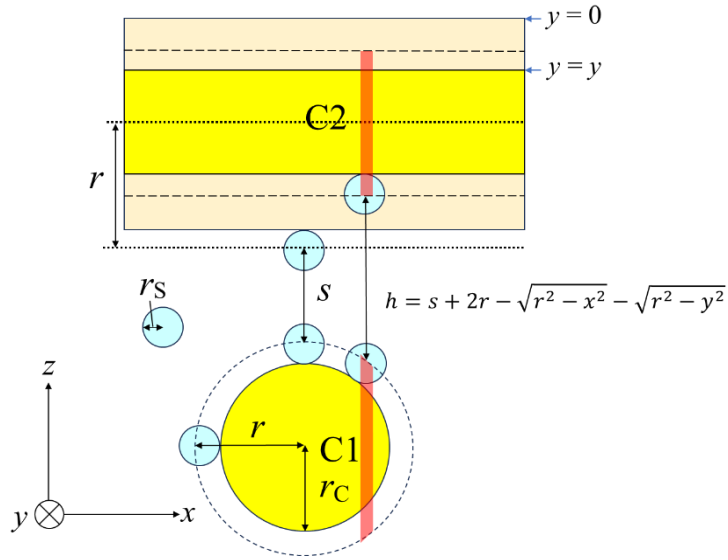

Fig .3: Illustration of the SFA system. Sum of the pressures between the red surface elements with distance $h$ corresponds to the force between the orthogonally arranged cylinders 1 and 2.

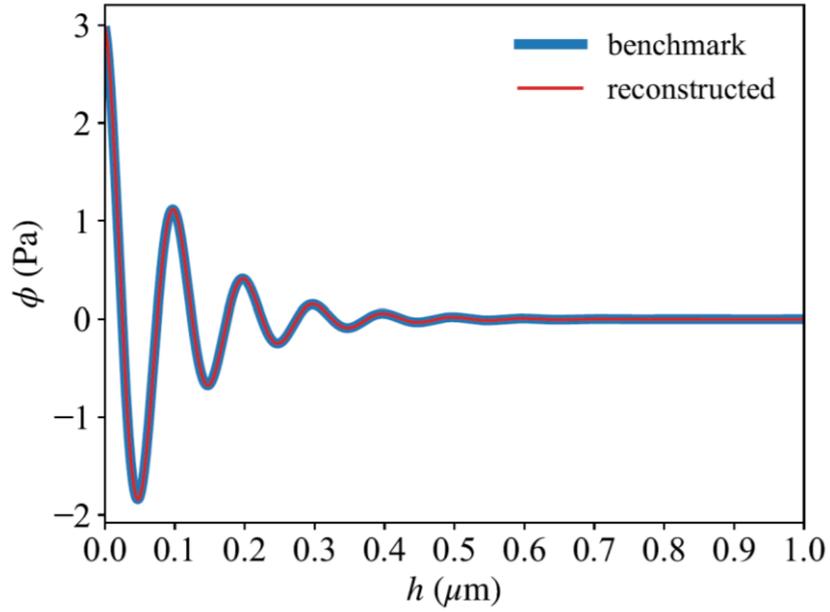

Fig .4: Demonstration of the PDA method in the SFA system.

**Conclusions.** — In this study, we introduced the PDA and δ-PDA methods and demonstrated them by applying integral equations in the OT, AFM, and SFA systems. The validity of these methods was analytically or numerically demonstrated. Overall, these methods provide a general and systematic framework for solving a wide range of inverse problems governed by integral equations. The simplicity and versatility of these methods make them valuable tools not only in surface science but also in other fields where integral equations play crucial roles. We expect that this approach will open new possibilities for various experimental and theoretical analyses.

***

We appreciate I. Ogasawara and M. Maebayashi for support of our research activity and environment. This research was supported by Tatematsu foundation.

*Data availability statement*: Data will be made available on request.

**REFERECES**

[1] VAN DEN BERG P. M. and KLEINMAN R. E., *Inverse Problems*, **13** (1997) 1607.

[2] LEVINSON H. W. and MARKEL V. A., *Phys. Rev. E*, **94** (2016) 43318.

[3] JAKOBSEN M., XIANG K. and VAN DONGEN K. W. A., *J. Acoust. Soc. Am.*, **153** (2023) 3151.

[4] PIJPERS F. P., *Mon. Not. R. Astron. Soc.*, **307** (1999) 659.

[5] GIESSIBL F. J., *Appl. Phys. Lett.*, **78** (2001) 123.

[6] SADER J. E. and JARVIS S. P., *Appl. Phys. Lett.*, **84** (2004) 1801.

[7] NISHI N., YAMAZAWA T., SAKKA T., HOTTA H., IKENO T., HANAOKA K. and TAKAHASHI H., *Langmuir*, **36** (2020) 10397.

[8] AMANO K. and TAKAHASHI O., *Physica A*, **425** (2015) 79.

[9] AMANO K., LIANG Y., MIYAZAWA K., KOBAYASHI K., HASHIMOTO K., FUKAMI K., NISHI N., SAKKA T., ONISHI H. and FUKUMA T., *Phys. Chem. Chem. Phys.*, **18** (2016) 15534.

[10] AMANO K., IWAKI M., HASHIMOTO K., FUKAMI K., NISHI N., TAKAHASHI O. and SAKKA T., *Langmuir*, **32** (2016) 11063.

[11] AMANO K., ISHIHARA T., HASHIMOTO K., ISHIDA N., FUKAMI K., NISHI N. and SAKKA T., *J. Phys. Chem. B*, **122** (2018) 4592.

[12] FURUKAWA S., AMANO K., ISHIHARA T., HASHIMOTO K., NISHI N., ONISHI H. and SAKKA T., *Chem. Phys. Lett.*, **734** (2019) 136705.

[13] HASHIMOTO K., AMANO K., NISHI N. and SAKKA T., *Chem. Phys. Lett.*, **754** (2020) 137666.

[14] OTAKE T., KAJITA R., OGASAWARA I., IWAKI M., ONISHI H., YOSHIMORI A. and AMANO K., *Physica A*, **647** (2024) 129918.

[15] OTAKE T. and AMANO K., *Phys. Lett. A*, **562** (2025) 131033.